%% file: 0-top.tex
\documentclass{sig-alternate-05-2015}
\usepackage{svg}
\usepackage{etoolbox}
\usepackage{tikz}
\makeatletter
\patchcmd{\maketitle}{\@copyrightspace}{}{}{}
\makeatother

\PassOptionsToPackage{bookmarks={false}}{hyperref}
\usepackage{soul}
\usepackage{pifont}
\usepackage{enumitem}
\usepackage{graphicx}
\usepackage{amsmath}
\usepackage{array}
\usepackage{mybeamer}
\usepackage{dirtytalk}

\usepackage[color]{changebar}
\usepackage{lipsum}
\newcommand{\nonl}{\renewcommand{\nl}{\let\nl\oldnl}}% Remove line for one line
\newcommand{\eg}{\mbox{{\em e.g.}}}

\newtheorem{definition}{Definition}[section]

\newrobustcmd*{\mysquare}[1]{\tikz{\filldraw[draw=#1,fill=#1] (0,0)
rectangle (0.2cm,0.2cm);}}

\newrobustcmd*{\mycircle}[1]{\tikz{\filldraw[draw=#1,fill=#1] (0,0) circle [radius=0.1cm];}}

\newrobustcmd*{\mytriangle}[1]{\tikz{\filldraw[draw=#1,fill=#1] (0,0) --
(0.2cm,0) -- (0.1cm,0.2cm);}}

\makeatletter
\newcommand{\removelatexerror}{\let\@latex@error\@gobble}
\makeatother

\usepackage{xcolor}

\begin{document}

\title{A Communication-Centric Observability Selection for Post-Silicon System-on-Chip Integration Debug}

\numberofauthors{2} %  in this sample file, there are a *total*
 \author{
% 1st. author
 \alignauthor 
 Yuting Cao, Hao Zheng \\
       \affaddr{CSE, U of South Florida, Tampa, FL}\\
       %\affaddr{Wallamaloo, New Zealand}\\
       \email{\{cao2, haozheng\}@mail.usf.edu}
 % 2nd. author
 \alignauthor
 Sandip Ray\\
       \affaddr{ECE, U of Florida, Gainesville, FL}\\
       %\affaddr{Dublin, Ohio 43017-6221}\\
       \email{sandip@ece.ufl.edu}
% % 3rd. author
% \alignauthor 
% Jin Yang\\
%       \affaddr{Strategic CAD Lab, Intel}\\
%       \affaddr{Hillsboro, OR}\\
%       %\affaddr{Hekla, Iceland}\\
%       \email{jin.yang@intel.com}
%\and  % use '\and' if you need 'another row' of author names
%% 4th. author
% \alignauthor Lawrence P. Leipuner\\
%       \affaddr{Brookhaven Laboratories}\\
%       \affaddr{Brookhaven National Lab}\\
%       \affaddr{P.O. Box 5000}\\
%       \email{lleipuner@researchlabs.org}
% % 5th. author
% \alignauthor Sean Fogarty\\
%       \affaddr{NASA Ames Research Center}\\
%       \affaddr{Moffett Field}\\
%       \affaddr{California 94035}\\
%       \email{fogartys@amesres.org}
%% 6th. author
%\alignauthor Charles Palmer\\
%       \affaddr{Palmer Research Laboratories}\\
%       \affaddr{8600 Datapoint Drive}\\
%       \affaddr{San Antonio, Texas 78229}\\
%       \email{cpalmer@prl.com}
}

\def\checkmark{\tikz\fill[scale=0.4](0,.35) -- (.25,0) -- (1,.7) -- (.25,.15) -- cycle;} 
\maketitle
\begin{abstract}
Reconstruction of how components communicate with each other during system execution  is crucial for debugging system-on-chip designs.  
However, limited observability is the major obstacle to the efficient and accurate reconstruction in the post-silicon validation stage.  This paper addresses that problem by proposing several communication event selection methods guided by system-level communication protocols. Such methods are optimized for on-chip communication event tracing infrastructure to enhance observability.  The effectiveness of these methods are demonstrated with experiments on a non-trivial multicore SoC prototype. The results show that with the proposed method, more comprehensive information on system internal execution can be inferred from traces under limited observability.

% This paper presents a post-silicon debug framework for SoC designs, with the specific objective of reconstructing system-level communication behavior from partially observed silicon traces.  Accurate reconstruction can offer SoC debuggers an abstract and structured view of the internal execution of system-level protocols, and facilitate the understanding and root-causing anomaly behavior observed during SoC debug.

%However, limited observability is the major obstacle to the efficient and accurate reconstruction.  That problem is addressed in this framework with a communication event selection method guided by system-level protocols, an on-chip communication monitoring infrastructure, and an off-chip trace analysis method specifically accounting for the system-level protocols.  This framework enhances observability, and enables efficient and accurate reconstruction of the internal executions for SoC designs. 
%We demonstrate the framework with  experiments on a non-trivial multicore SoC prototype  and further show that the on-chip  monitoring infrastructure incurs very little overhead in area and logic complexity. 
\end{abstract}

\footnotetext{\textcopyright 2019 IEEE.  Personal use of this material is permitted.  Permission from IEEE must be obtained for all other uses, in any current or future media, including reprinting/republishing this material for advertising or promotional purposes, creating new collective works, for resale or redistribution to servers or lists, or reuse of any copyrighted component of this work in other works.}

\input{1-intro}
\input{2-background}

\input{3-monitor}

% \input{3-analysis}
%\input{5-gap}
\input{4-selection}
\input{5-experiments}

\input{6-rel-work}

%************
\section{Conclusion}

This paper describes several communication event selection methods aiming to boost flow execution coverages under limited observability.  The observed traces on the selected communication events capture more comprehensive information to allow better understanding of the communications of components during system execution, thus facilitating debug.    In the future, we plan to perform in-depth studies of the proposed methods on more complex SoC designs with diverse interconnects such as the RocketChip SoC design. 

\vspace{3pt}
\noindent{\bf Acknowledgment~}
The work presented in this paper is partially supported by a gift from the Intel Corporation. 
%The authors would like to thank anonymous reviewers for their constructive comments.

%\bibliographystyle{unsrt}
%\bibliography{soc.bib}  

\input{0-top.bbl}
\end{document}

%% file: 1-intro.tex
\section{Introduction}
\label{sec:intro}
% \subsection{Motivation}

% An execution trace of a system typically involves activities from
% the CPU, audio controller, display controller, wireless
% radio antenna, etc., reflecting the interleaved execution of a potentially large number of 
% communication protocols. 
%For example, consider a smartphone executing a usage scenario where the end-user browses the Web while listening to music and sending and receiving occasional text messages.  Typical post-silicon validation use-cases involve exercising such scenarios.  

A modern System-on-Chip (SoC) design is typically constructed by composing a large number of pre-designed 
% hardware or
% software blocks often referred to as 
\say{intellectual properties} or \say{IPs} that coordinate through complex
protocols to implement system-level behavior.%~\cite{Foster2015DAC}.  
Over the last decade, the number and heterogeneity of IPs integrated in an SoC design have continued to grow, and the trend is towards an even sharper growth gradient as we develop sophisticated systems targeting complex applications like automotives and Internet-of-Things.  Unsurprisingly, this has led to an increasing count of design bugs \cite{yerramili,Patra2007}.  These systems are often deployed in critical applications, bugs discovered after their deployment in field can be extremely expensive, resulting in catastrophic loss of company revenues,  compromise of personal and national security, and even human life.

Post-silicon debug is a critical component of the validation of modern microprocessors and SoC designs.  It is performed on silicon implementation, to make sure that the finished system works as intended under actual operating conditions.
Unfortunately, it is also highly complex, performed under aggressive schedules and accounting for more than $50\%$ of the overall design validation cost~\cite{Patra2007}.
A major challenge in post-silicon debug is the severely limited observability where only a small number of debug interface signals are available to observe a vast space of internal executions of SoC designs.

In this paper, we consider post-silicon {\em integration debug} of SoC designs, which concerns  debugging anomalies in executions of communication protocols among various IPs.  
% Errors in these protocols form some of the most subtle and hard-to-debug errors in validation.  This is because, 
Executions in modern SoC designs entail significant interleavings of a large number of such protocols, and errors can occur because of a subtle race condition in a specific interleaving which is difficult to exercise or repeat~\cite{Talupur:2015:fmcad}.  Furthermore, while individual IPs are often reused across products, %(resulting in their core functionality being hardened through multiple validation across products) 
their specific integration, --- and consequently, the protocols involved in their communications ---, is unique to each individual SoC.   This results in unique bugs arising in the communication component of each SoC, which are hard to isolate, replay, triage, and root-cause.  The situation is particularly exacerbated by the fact that due to partial observability only a small set of events in the participating protocols can be actually observed in each execution, making it harder to pinpoint the exact interleaving involved in the execution.

In order to address the above challenge, this paper presents a number of communication event selection methods aiming to increase coverage metrics that are relevant to various debug objectives.  These methods are optimized for an on-chip communication event tracing infrastructure that can be typically found in modern SoC designs.   The main {\bf contributions} of this paper include the following.
\begin{itemize}[leftmargin=*]
\setlength{\itemsep}{0pt}
    \item This work considers enhancing observability with respect to an on-chip real-time tracing infrastructure instead of trace buffers.  Using trace buffers can only offer a limited window of observability while the real-time tracing provides more comprehensive observability over the entire course of system execution, which is indispensable for SoC integration debug.
    \item New coverage metrics are proposed for evaluating relevance and comprehensiveness of information captured on observed traces with respect to system-level communication protocols.  The traditional metric, {\em state restoration ratio}~(SRR), is not applicable for SoC integration debug. 
    \item Communication event selection methods are driven by the proposed coverage metrics optimized for the real-time tracing infrastructure so that observed traces only capture the most relevant communication events under limited observability.
\end{itemize}

%% file: 2-background.tex
\section{Background}
\label{flow-spec}
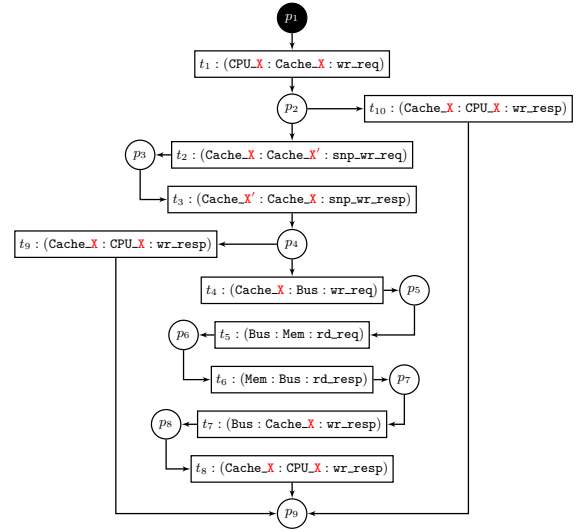
\begin{figure}[tb]
\begin{center}
\resizebox{.43\textwidth}{!}{
\input{lpn-flow.tex}
}
\caption{Graphical representation of a CPU write flow.}
\label{fig:ex}
\end{center}
\vspace{-5mm}
\end{figure}

% \cbcolor{pink}
% \cbstart
In architectural documents, system-level protocols are often represented as message flows, therefore they are  referred to as system flows in this paper. 
\begin{definition} A \textbf{system flow} is defined as a tuple {${F} = (P, T, E, L)$} where {$P$} is a finite set of \emph{places}, $T$ is a finite set of \emph{transitions}, $E$ is a finite set of \emph{events}, and $L: T \rightarrow E$ is a labeling function that maps each transition $t \in T$ to an event $e \in E$.
\end{definition}
An SoC typically implements several flows denoted as $\vec{F}$.  In this paper, $F_i \in \vec{F}$ denotes one such flow.
In a system flow, an event is a tuple $({\tt src, dest, cmd})$ where ${\tt cmd}$ is a command sent from a source component ${\tt src}$ to a destination component ${\tt dest}$.  An event is generated when block {\tt src} communicates with block {\tt dest}.  An example CPU write flow is shown in Figure~\ref{fig:ex}.

For each transition $t \in T$, its preset, denoted as $\bullet{t} \subseteq P$, is the set of places connected to $t$, and its postset, denoted as $t\bullet \subseteq P$, is the set of places that $t$ is connected to.  A state $s \subseteq P$ of a flow is a subset of places marked with tokens.   There are two special states associated with each flow; 
$s_0 \subseteq P$ is the set of initially
marked states, also referred to as the \emph{initial state}, and the end state $s_{\perp} \subseteq P$ is the set of end states not going to any transitions.  Each flow is associated with one \emph{start} and several \emph{end} events.  An event $e \in E$ is a start event if $e = L(t)$ and $\bullet t \subseteq s_0$.  An event $e \in E$ is an end event if $e = L(t)$ and $t \bullet \subseteq s_{\perp}$. In Figure~\ref{fig:ex}, $s_0 = \{p_1\}$, and $s_{\perp} = \{p_9\}$, its start event is {\tt (CPU\_X:Cache\_X:wr\_req)}, and its end event is the one labeled for transitions $t_8, t_9$ and $t_{10}$.  The occurrence of a start event indicates the beginning of a flow execution, while the occurrence of an end event indicates the complete of a flow execution. 

A transition $t$ can be fired in a state $s$ if $\bullet t \subseteq s$.  Firing $t$ causes the labeled event to be emitted, and leads to a new state $s^\prime = (s - \bullet t) \cup t \bullet$. Therefore, executing a flow induces a sequence of events.  Execution of a flow completes if its $s_{\perp}$ is reached.  
% For example, in Figure~\ref{fig:ex}, $t_1$ can be executed in $s_0 = \{p_1\}$.  Event ${\tt (CPU\_X:Cache\_X:wr\_req)}$ is emitted after $t_1$ is executed, and the LPN state becomes $\{p_2\}$.  The  end state is $s_{end} =\{p_{9}\}$. 

%A flow specification may contain multiple branches describing different ways a system can execute such flow. For example, the flow shown in Figure~\ref{fig:ex} has three branches covering the cases where the cache snoop operation is hit or miss.

\begin{definition}
An {\bf instance} of a flow $F_i \in \vec{F}$ is $F_{i,j}$ where $j$ denotes the instance index.
Similarly, every element of $F_{i,j}$ is an instance of the corresponding element in $F_i$.
\end{definition}

During an execution of a SoC design, instances of the set of flows $\vec{F}$ that it implements are executed.  Suppose that the set of flow instances executed is $\{F_{i,j}~|~F_i \in \vec{F}\}$.  The following definition relates flow executions with event sequences.

\begin{definition} Given a SoC design that implements a set of flows $\vec{F}$, an \textbf{execution} on its instances $\{F_{i,j}~|~F_i \in \vec{F}\}$ yields a \textbf{trace $\rho$} such that 
\[
\rho = \{e_0 e_1 \ldots e_n\ |\ e_i \in \bigcup E_{i,j} \}
\]
{where $E_{i,j}$ is the set of events of flow instance $F_{i,j}$. }
\end{definition}
From a trace, an execution of flow instances can be inferred by following the transition firing semantics of system flows defined above.

% \textbf{Definition 4.}We use \textbf{Flow execution scenario} to represent  information that can be extracted from a given flow trace $\rho$. It can be viewed as \emph{a state of system execution abstracted wrt system flows}, and it is defined as a set of states of individual flow instances.
% \[ 
% scen = \{ (F_{i,j}, \mathit{start}_{i,j}, \mathit{end}_{i,j}, s_{i,j})~|~F_i \in {\bf F}\}
% \]
% Here each flow instance's entry includes its identity $F_{i,j}$,   $\mathit{start}_{i,j}$ and $\mathit{end}_{i,j}$ representing relative time when $F_{i,j}$ is initiated and completed, and $s_{i,j}$ that is the current state of $F_{i,j}$. 
% The ordering relations among flow instances can be derived by comparing their {\em start} and {\em end} indices. 
% For example, for two flow instances in an execution scenario, $(F_{u,v}, \mathit{start}_{u,v}, \mathit{end}_{u,v})$ and 
% $(F_{x,y}, \mathit{start}_{x,y}$, $\mathit{end}_{x,y})$,  $F_{u,v}$ is initiated before $F_{x,y}$ if $\mathit{start}_{u,v} < \mathit{start}_{x,y}$, or $F_{x,y}$ is initiated after $F_{u,v}$ is completed if $\mathit{end}_{u,v} < \mathit{start}_{x,y}$. The ordering relations can provide more accurate information for understanding system execution under limited observability.  

%% file: lpn-flow.tex
\begin{tikzpicture}[node distance=3cm, auto,>=latex', thick]
\tikzset{rectbox/.style={rectangle, draw, align=flush center,thick, minimum size = 6mm}}
\tikzset{circ/.style={circle, draw,thick}}
\tikzset{terminal/.style={circle, draw,line width=.8mm}}
\tikzset{weight3/.style={line width=.3mm}}
	% Nodes:
	\node[circ,fill=black]		(p1) 		at (0,0) {\textcolor{white}{$p_1$}};
	\node[rectbox]	(t1)	at	($(p1.south) + (0,-0.7)$) {$t_1:( {\tt CPU\_{\color{red}X}: Cache\_{\color{red}X}: wr\_req})$};
	\node[circ] 	(p2) 		at ($(t1.south) + (0,-0.7)$) {$p_2$};
	\node[rectbox]	(t2)	at	($(p2.south) + (0,-0.7)$) {$t_2:( {\tt Cache\_{\color{red}X}: Cache\_{\color{red}X'}: snp\_wr\_req} )$};
	\node[circ] 	(p3) 		at ($(t2.west) + (-0.7,0)$) {$p_3$};
	\node[rectbox]	(t3)	at	($(t2.south) + (0,-0.7)$) {$t_3:( {\tt Cache\_{\color{red}X'}:  Cache\_{\color{red}X}: snp\_wr\_resp})$};
	\node[circ] 	(p4) 		at ($(t3.south) + (0,-0.7)$) {$p_4$};
    \node[rectbox]	(t4)	at	($(p4.south) + (0,-0.7)$) {$t_4:({\tt Cache\_{\color{red}X}: Bus: wr\_req})$};
    \node[circ] 	(p5) 		at ($(t4.east)+(00.7,0)$) {$p_5$};
	\node[rectbox]	(t5)	at	($(t4.south) + (0,-0.7)$) {$t_5:( {\tt Bus: Mem: rd\_req} )$};
    \node[circ] 	(p6) 		at ($(t5.west) + (-.7,0)$) {$p_6$};
    \node[rectbox]	(t6)	at	($(t5.south) + (0,-0.7)$) {$t_6:( {\tt Mem: Bus: rd\_resp} )$};
    \node[circ] 	(p7) 		at ($(t6.east) + (0.7,0)$) {$p_7$};
	\node[rectbox]	(t7)	at	($(t6.south) + (0,-0.7)$) {$t_7:( {\tt Bus:Cache\_{\color{red}X}:wr\_resp})$};
    \node[circ] 	(p8) 		at ($(t7.west) + (-0.7,0)$) {$p_{8}$};
   \node[rectbox]	(t8)	at	($(t7.south) + (0,-0.7)$) {$t_{8}:( {\tt Cache\_{\color{red}X}: CPU\_{\color{red}X}: wr\_resp} )$};
   \node[rectbox]	(t9)	at 	($(p4) + (-4,0)$) {$t_{9}:( {\tt Cache\_{\color{red}X}: CPU\_{\color{red}X}: wr\_resp} )$};
   \node[rectbox]	(t10)	at 	($(p2) + (4,0)$) {$t_{10}:( {\tt Cache\_{\color{red}X}: CPU\_{\color{red}X}: wr\_resp} )$};
	\node[circ] 	(term1) 		at ($(t8.south) + (0,-0.7)$) {$p_{9}$};
	
	% Edges
 	\draw[->,weight3]	(p1) -- (t1);
 	\draw[->,weight3]	(t1) -- (p2);
 	\draw[->,weight3]	(p2) -- (t2);
 	\draw[->,weight3]	(t2) -- (p3);
 	\draw[->,weight3]	(p3) |- (t3);
 	\draw[->,weight3]	(t3) -- (p4);
 	\draw[->,weight3]	(p4) -- (t4);
 	\draw[->,weight3]	(t4) -- (p5);
 	\draw[->,weight3]	(p5) |- (t5);
 	\draw[->,weight3]	(t5) -- (p6);
 	\draw[->,weight3]	(p6) |- (t6);
 	\draw[->,weight3]	(t6) -- (p7);
 	\draw[->,weight3]	(p7) |- (t7);
 	\draw[->,weight3]	(t7) -- (p8);
   	\draw[->,weight3]	(p8) |- (t8);
	\draw[->,weight3]	(p2) -- (t10);
    \draw[->,weight3]   (t10) |- (term1);
	\draw[->,weight3]	(p4) --  (t9);
	\draw[->,weight3]	(t9) |- (term1);
 	\draw[->,weight3]	(t8) -- (term1);
\end{tikzpicture}

%% file: 3-monitor.tex
\section{Communication Tracing}
\label{m-infra}

\begin{figure}[tb]
\begin{center}
\includegraphics[width=\linewidth]{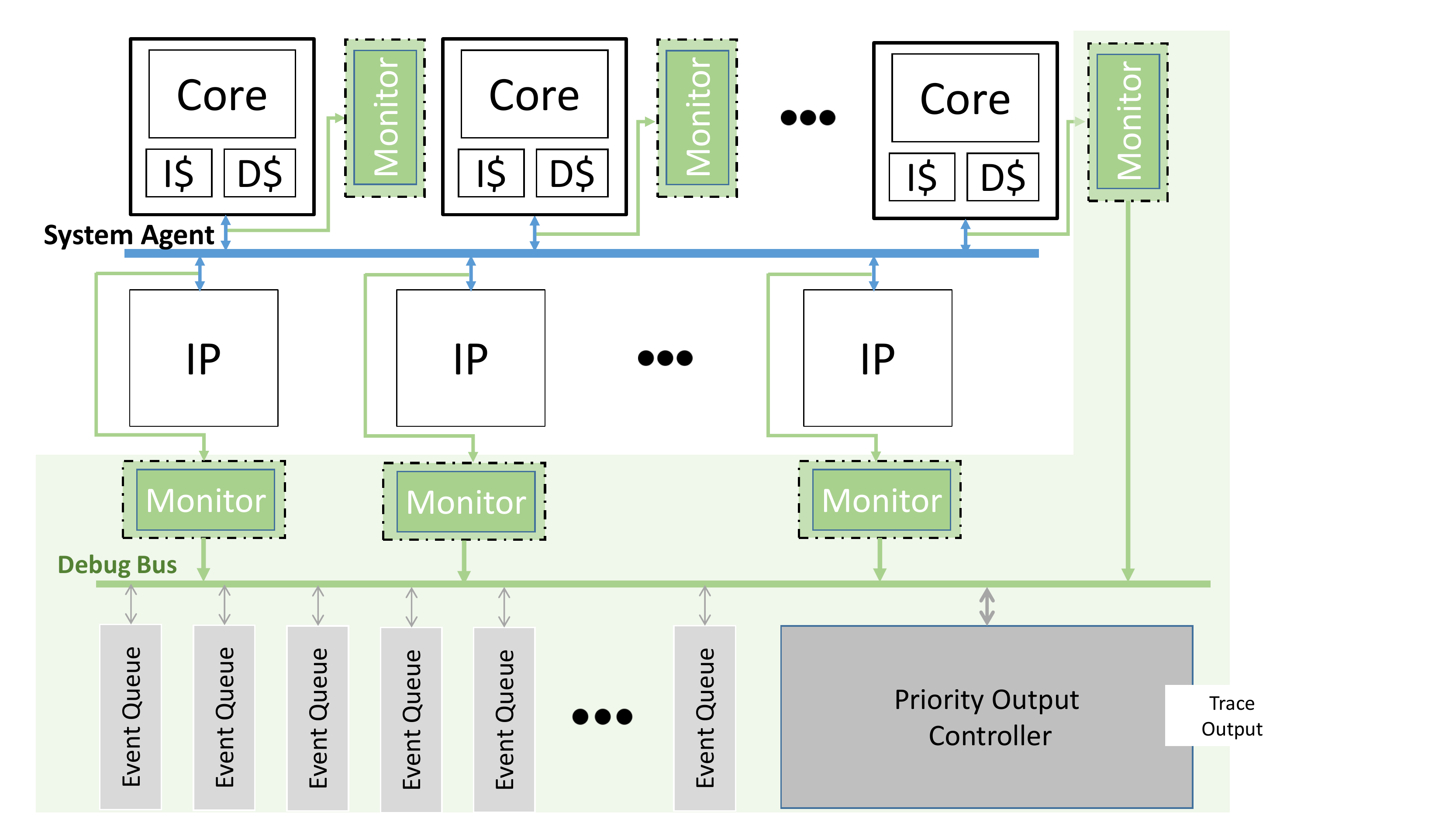}
\caption{Architecture of the Communication Tracing Module (CTM) for SoC integration debug.}
\label{fig:funnel2}
\end{center}
\vspace{-3.0mm}
\end{figure}

System execution typically involves a number of flow instances executed concurrently.  In order to reconstruct flow executions off-chip, communication events involved in flow executions must be collected and off-loaded.  Modern SoC designs are typically instrumented with extensive dfx circuitry for various debug functions.  This section describes an on-chip infrastructure for real-time tracing of communication events during system execution.   The architecture of this infrastructure is shown in Figure~\ref{fig:funnel2}.

%
%*****************
\subsection{Communication Tracing Module (CTM)}
\label{sec:monitor}

% \begin{figure}[tb]
% \centering{
% \includegraphics[width=.4\textwidth]{figures/axi_read.png}
% }
% \caption{AXI read protocol~\cite{AXI4}.}
% \label{axi_read_fsm}
% \end{figure}

% \begin{figure}[tb]
% \centering{
% \includegraphics[width=.8\linewidth]{figures/wavedrom}}
% \caption{An example of the AXI read transaction on a communication link, and the output of a monitor attached to that link.}
% \label{fig:waveform}
% \end{figure}

The communication tracing module consists of communication monitors and an off-load unit.  Communication monitors are attached to communication links of interest.  A monitor observes signal events transferred on a link, and generates a communication event if a pattern on the observed signal events is recognized.  
A \emph{signal event} is an assignment to a set of design signals. A \emph{flow event} is an abstract construct used in flow specifications, and is typically implemented by a sequence of signal events.  Communication events generated by monitors can be viewed as encodings of flow events in a SoC design.
% Figure~\ref{fig:waveform} show the waveform where a master reads a slave through the AXI read protocol~\cite{AXI4}.  The blue lines (top 3 lines) are design signals as inputs to the attached monitor, while the red lines (bottom 2 lines) are the outputs of the monitor.  From this example, it can be seen that 
An important function of a monitor is to compress a sequence of signal events across a potentially large number of cycles into a single cycle communication event, which is beneficial to reduce the bandwidth demand on the trace interface. 
Once a communication event is detected, a monitor can selectively encode information that is useful for the off-chip analysis including operation commands, addresses, etc. 
%The basic idea of the above monitor can be naturally extended to different protocols such as the AXI write request and response. 

% \subsection{Transaction Monitors}
% \begin{figure}[t]
% \centering{
% \includegraphics[width=.45\textwidth]{figures/monitor.png}
% }
% \caption{Monitor is inserted between AXI master and slave to observe all communication signals.}
% \label{monitor}
% \end{figure}

% It is possible that during debug only a subset of flows are observed either because they are necessary for understanding a particular use case or due to the limited observability of the trace interface.   This requires to limit flow events to observe.  Therefore, the monitors in our framework is designed such that they can be configured to output events that meet certain features.
% Furthermore, the monitors can also detect low level protocol errors timely and right on the spot.
% For example, the monitor shown in Figure~\ref{fig:waveform} can output an unique {\tt ERROR} event if {\tt M\_Read\_Val} is set without {\tt S\_Read\_Ready} set in the previous cycle.

%
%*************
% \subsection{Communication Event Output}
% \label{sec:funnel}

Communication events from  monitors can be stored in the on-chip trace buffers, and off-loaded at the end of system execution.  However, the on-chip trace buffers can only store a limited number of events due to their limited capacities. 
On the other hand, our communication tracing infrastructure includes an event off-load unit that can off-load events via trace ports on-the-fly, thus enabling system internal executions over an much extended period to be observed for off-chip analysis. 

Since communcation events from monitors are typically distributed over time relatively sparsely, the off-load unit interleaves events, and off-loads them via the shared trace ports in a time-multiplexing manner.   An issue with the interleaved approach is that the rate of events detected by monitors can exceed the peak bandwidth of the trace ports.  In that case, certain events have to be dropped.  The inability to off-load all events occurred during an execution can be viewed as another form of limited observability.  In order to reduce the number of events that have to be discarded, output of each monitor is connected a queue that buffers events waiting for off-load.  On every cycle, events from all monitors are stored into the corresponding queues. 
An output controller scans all event queues, and off-loads buffered events one at a time.
% As shown in Figure~\ref{fig:funnel2}, all monitors are connected to the event output unit.  Events from the monitors are routed through this unit, merged into a sequence, and  eventually output through the trace port.  The biggest advantage  may be the potentially very high observability in terms of a large number of communication links to be traced and the large amount of information that can be encoded with all available trace signals for each individual event.   

\subsection{Limited Observability}
\label{sec:priority-output}

As indicated in the previous section, limited bandwidth of trace ports may cause some communication events not to be observed.  The frequency of event dropping is roughly affected by the gap between the rate of events generated by all monitors and the bandwidth of trace ports.  The factors affecting the rate of communication event generation include the design micro-architecture, test programs used during debug, and the number of flows and flow events selected for observation.  One simple technique to reduce the frequency of event dropping is large buffers for events waiting for output, but it leads to a large area overhead. 

% With the understanding that event dropping is almost inevitable, this paper presents a technique with a goal of reducing the impact of event droppings on the accuracy and efficiency of the trace analysis.  
% To minimize the effect of event dropping that can be hardly avoided, 
An important observation exploited in this paper is that some events are more important for understanding flow executions than others. If the less important events are not observed, then the whole trace port bandwidth can be dedicated to observing the more important events.  Furthermore,  not observing certain events may eliminate the need of observing certain communication links altogether.  In that case, the corresponding monitors can be disabled, and the capacities of the associated event queues can be re-allocated to the queues for the links under observation.
By increasing the queue capacities for the links under observation, the event droppings can be effectively reduced.   This hypothesis about the relation between the number of links to observe and the event losses will be exploited and experimentally validated in the following sections.  %Although this enhancement is achieved at the cost of fewer observation of less important events, it provides a more comprehensive understanding of the system behavior. 
Next section presents metrics for evaluating importance of events, and methods for selecting important events to observe.

% observing a smaller set of important events effectively improve the observability of the system beh }
% Therefore, it is preferable to drop events that are less important. 
% During an execution run, the event output unit can be configured to enter the \textbf{priority output mode (POM)} under certain conditions where events with higher priorities are outputted first. %If any non-critical events are observed under priority output mode, it is directly dropped.

%For example, compared to other transaction event happens during inside a flow instance, its initialization event is more important because if the initialization event of a flow event is dropped, the trace analysis conducted off-chip will not be able to obtain an timing order among flow instances. 

% The priority output mode is controlled on-the-fly during system execution by an output configuration register~({\tt OCR}), which has the number of bits equal to the number of monitors.  This register is used as a mask for the output from {\tt Event Validity Queue}.  In the normal mode, all bits in {\tt OCR} are set to $1$.  Suppose that the threshold for link $i$ is crossed at some point,  the output unit enters the priority output mode by setting the $jth$ bit of {\tt OCR} to $0$ for each link $j$ that is ranked lower than link $i$.  This would effectively block events from the $jth$ monitor from being output.

%% file: 4-selection.tex
%
%**************
\section{Flow Event Selection}
\label{sec:observe-config}

% A crucial activity during post-silicon debug is to identify flow instances involved in an execution from an observed trace.  In the rest of this paper, this activity is referred to as {\em flow execution inference}.
% As discussed in Section~\ref{sec:priority-output}, limited bandwidth of the trace port and capacity of on-chip buffers inevitably cause event losses.  This may cause missing important information regarding what flow instances are executed {\em wrt} an observed trace.  

As shown in the previous section, not observing less important events can help prevent important events from being dropped.  
In this section, importance of events is characterized by a number of coverage metrics for different debug purposes.  After defining the coverage metrics, we describe communication event selection methods driven by those coverage metrics.  The objective of these selection methods is to maximize the coverage metrics under the resource constraints of the communication tracing infrastructure. 
Unlike previous work on trace signal selection~\cite{mishra2011vlsi,Ma2015ICCAD}, the methods presented in this section target selecting flow events for observation instead of individual design signals. 

% \subsection{Debugging Process}
% \label{sec:debug-process}
% \textcolor{blue}{
% ---------------Not sure where to put this section------------
% Post silicon debugging is a broad to narrow process. 
% In the early stage, while not much information is available, it is important to obtain the overall picture of all flows activated (types and number of each flow). Thus the flow event selection is mainly targeted at providing comprehensive interpretation of the system behavior, and helping the debugger to make logical projection about the suspicious IPs. As the information regarding the buggy behaviors becomes more complete, the debugging scope is narrowed down to certain suspicious areas.
% % revealing more details of the related system behavior. 
% At each iteration, information of the previous rounds are combined to assist the new selection. Such combined information includes the activated flow instances, their relative orderings and suspicious flows. Depending on the main goal of the current iteration, the flow event selection is guided by different metrics, discussed in next section.
% }

\subsection{Selecting Flows to Observe}

An critical activity performed in post-silicon stage is to validate application usage scenarios.  In this activity, individual target usage scenarios, \eg, for a smartphone, playing videos or surfing the Web, while receiving a phone call, are exercised, while possible failures \eg, hangs, crashes, deadlocks, overflows, etc., are monitored.   Usage scenario validation forms a key part of SoC integration validation.  Each usage scenario usually involves interleaved execution of several flows among IPs in the SoC design, \eg., a usage scenario that entails receiving a phone call in a smartphone when the phone is asleep involves flows among the antenna, power management unit, CPU, etc.  Therefore, only the flow events of the involved flows in a usage scenarios are typically observed.  
 As explained in the previous section,  observing a restricted subset of events can reduce incidients of events being dropped. 

\subsection{Flow Execution Coverage Metrics}

In this section, we consider the problem of characterizing the important or relevance of flow events by defining coverage metrics for different debug purposes.  

The first metric is \textbf{flow instance coverage (FIC)}, which is defined below.  
\begin{equation} \label{eq:FIC-scen}
FIC = \frac{I}{N}
\end{equation}
where $I$ is the number of flow instances observed, and $N$ is the total number of flow instances executed.  FIC defines a fraction of the number of flow instances actually observed versus the total number of flow instances executed.   We say that a flow instance is observed in a trace if any event of that flow instance is observed in the trace.   Note that the parameter $N$ is not essential when this metric is applied to evaluate different observabilities as all of them are evaluated assuming the same $N$.  

The purpose of FIC is to offer a metric to evaluate different observabilities to support a coarse-grained global view of system execution. In this coarse-grained global view, we are interested in all flow instances executed in the entire course of a debug run, instead of detailed execution of individual flow instances.   It may provide valuable information about anomaly behavior in system execution, \eg, an unusually high number of wakeup calls to a CPU from the power management unit.   In this case, we want to select an observability that maximize FIC.  Obviously, one observability is better than another one if its FIC derived from an observed trace is higher. 

\begin{figure}[tb]
\begin{center}
\resizebox{2.8in}{!}{
\includegraphics[width=.5\textwidth]{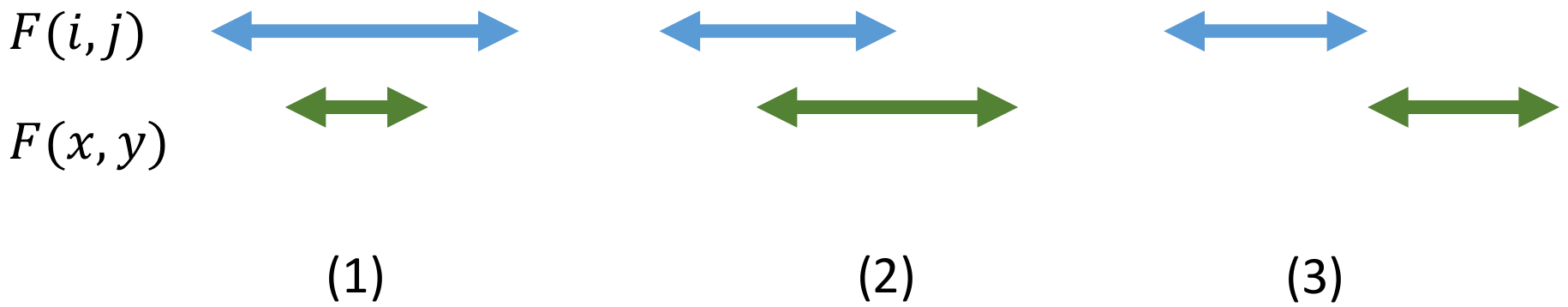}
}
\vspace{-1.5mm}
\caption{Possible interleavings between two flows}
\label{fig:interleave}
\end{center}
\vspace{-6.5mm}
\end{figure}

As explained in the introduction, many errors are due to intricate interleaved execution of flows, \eg, a firmware execution flow executed before a firmware authenticate flow completes, signaling a security breach.  Therefore, the coarse-grained global view obtained by only observing flow instances is inadequate.  Observed traces must capture sufficient information to allow interleaved execution of different flow instances to be extracted.   The interleaving relations between two flow instances $F_{i,j}$ and $F_{x,y}$ are shown in Figure~\ref{fig:interleave}.  In this figure, the length of arrows shows the duration of a flow instance execution, while the arrows at both ends indicate the time when a flow instance is initiated and when it completes.   Figure~\ref{fig:interleave} shows three possible interleavings: (1)  $F_{i,j}$ starts before $F_{x,y}$ starts, and it completes after $F_{x,y}$ completes; (2)  the initiation and completion of $F_{i,j}$ occur earlier that the initiation and completion of $F_{x,y}$; (3) $F_{i,j}$ completes before $F_{x,y}$ starts. 
An observability needs to be selected in order to support such interleavings to be captured on the observed traces.
To evaluate different observabilities, we define the \textbf{complete execution coverage (CEC)} metric as below.
\begin{equation} \label{eq:CEC}
CEC  = \frac{C}{N}
\end{equation}
where $C$ is the number of complete flow instances extracted from an observed trace, and similarly $N$ is the total number of flow instances executed.  A {complete} flow instance is observed if both its start and end events are found in the observed trace.

\subsection{Coverage Driven Event Selection}
\label{sec:event_selection}

This section presents observability selection methods driven by coverage metrics described in the previous section.   The inputs to a selection method  are a set of flows to observe and a coverage metric, and the output is a subset of flow events for observability that maximize the coverage metric. 
% To achieve the test performance considering the two metrics in the previous section, each flow event is ranked with a priority value, and the rank of a communication link is the same as the highest rank of an event transferred over that link.
% When selecting sets of events for evaluation, it is unpractical to take all combinations. This is because the numbers of flow events can be big. Therefore, it is helpful to consider the important information encoded in the events.

To select an observability targeting coverage metric FIC, it is necessary to select a subset of flow events that cover all flows under observation such that an event in an observed trace can uniquely identify a flow instance.  In practice, most SoC designs include architectural support for tagging, which allows uniquely identifying different flow instances from observing properly tagged events. 
Because of the unique correspondence between flow instances and observed events, all flow events are FIC-equivalent.  Therefore, we aim to select a subset of events that maximizes the FIC.  

Given a set of flows to observe, there can be many different selections of events of those flows.  In order for the observed traces on the selected events to have high FIC, the losses of the selected events must be low.  Recall the hypothesis presented in the previous section indicating that the losses of events can be reduced if the number of links under observation is reduced.  By this hypothesis, for two sets of selected events, if the number of links to observe for one set of selected events is smaller than that of another set, then the former is preferred. 

% each flow event can be uniquely mapped to a flow instance. The observation of any flow event indicates the corresponding flow is exist, increasing the overall $FIC$.
% Whereas, the $FIC$ stays the same even if more than one flow events are observed in a flow instance.
% Therefore, with limited on-chip resources, we try to make sure at least one flow event is selected for each flow to maximize $FIC$. 

Next, we consider observability selection targeting the coverage metric CEC.  In order for observed traces to have high CEC, the start and end events of all flow instances should be observable.  Therefore, the start and end events of all flows to observe must be selected.   
% This CEC is highly depended on the amount of critical events observed as it contains the essential information of the complete flow instances. All non-critical events are CEC-equivalent. 

To facilitate more effective debug, it is necessary to know additional information beyond the initiation and completion of each flow instance.  More specifically, it would be useful to know which path in the flow is followed when an instance of that flow is executed.  Consider the flow in Figure~\ref{fig:ex} fpr an example. It has three possible execution paths.  Observing only the start and end events (labelings of $t_1$ and $t_{10}$) is not sufficient to tell which execution path is actually followed.  Given an observed trace shown below, 
\begin{center}
    {\tt (CPU\_x:Cache\_x:wr\_req)}, \ {\tt (Cache\_x:CPU\_x:wr\_resp)}, \ \ldots
\end{center}
we are not able to confirm whether it is a result of executing that flow following the rightmost path ($p_1$, $t_1$, $p_2$, $t_{10}$, $p_9$) or one of the other two paths with all the intermediate events not observed.  To obtain the information on paths following by a flow execution, some unique event from each path needs to be selected.  Consider the same flow example.  Either one of $\{t_2, t_3\}$ needs to be selected in order to identify the leftmost or middle path.  Moreover,  one event from $\{t_4, t_5, t_6, t_7\}$ needs to be selected to identify the middle path.  As the above illustration shows, there are different ways to select additional events to observe for detailed executions of a set of flows.  Similarly, these different selections are evaluated based on the number of links that need to be observed for the selected events. 

\cbend

%% file: 5-experiments.tex
\section{Experiments}
\label{section:results}

%To the best of our knowledge, this work is the first to present a system-level post-silicon debug framework for SoC designs guided by system level protocols.  We are not able to find any similar previous work where ours can be evaluated and compared with.  The closest work to ours is \cite{Zheng2016ISQED}.  However, our work is more general and developed with practical considerations.  Additionally, the work in~\cite{Zheng2016ISQED} is discussed and evaluated based on an abstract event-level model while our approach is evaluated on an RTL model.

%\red{\em Need to emphasize that difficulty to obtain the code for existing signal selection methods and to compare with signal selection proposed in this paper}

%--
\subsection{The Model}

% Due to the simulator used for experiments (GHDL~\cite{ghdl}, an open-source simulator), the memory size cannot be large, even though the system bus is 32-bit wide. 
\begin{figure}
\begin{center}
\resizebox{3.4in}{!}{
\includegraphics[width=\linewidth]{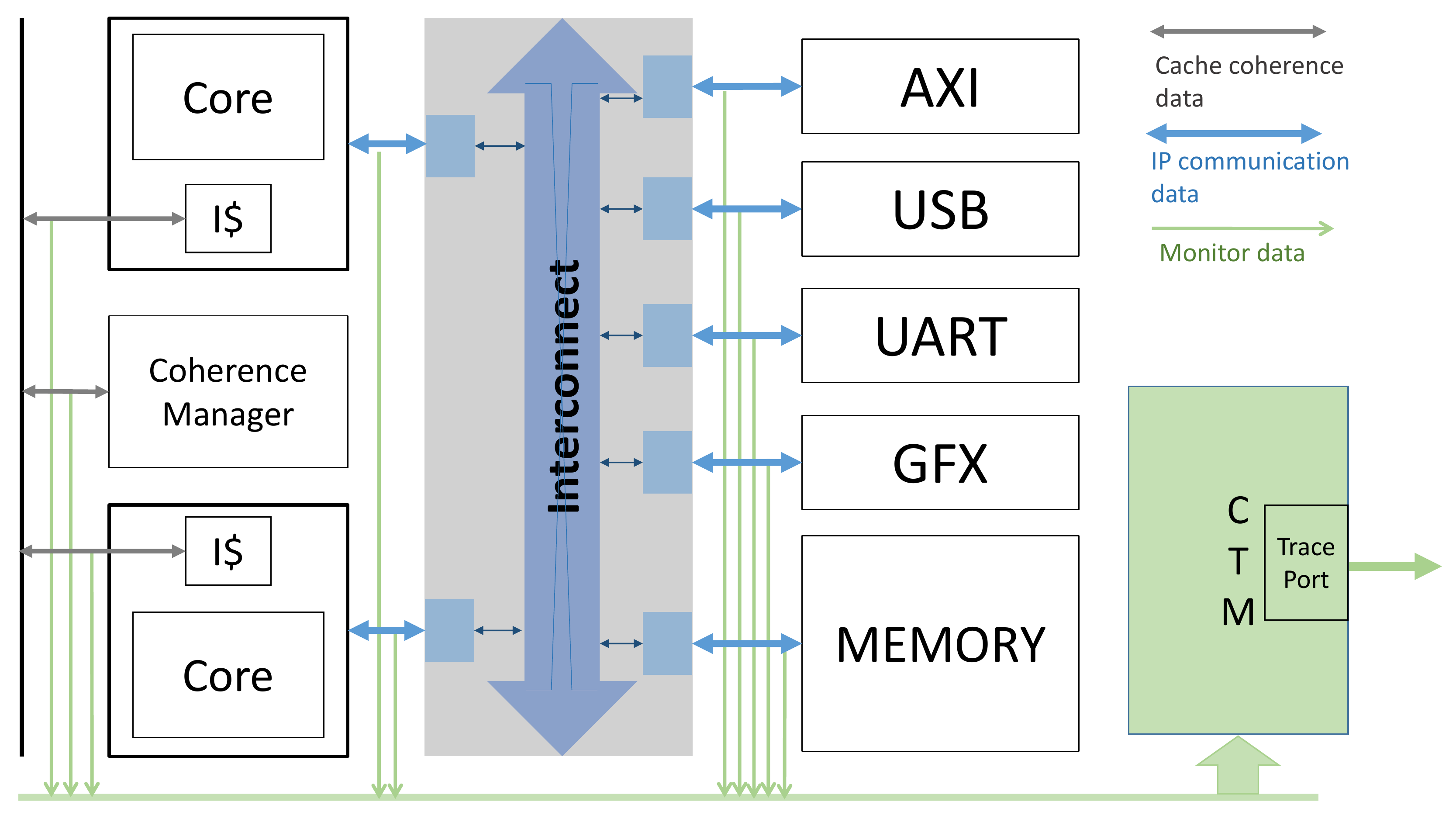}
}
\caption{A SoC prototype where each communication link is attached with a monitor. There can be multiple links between a pair of components.}
\label{rtlstruc}
\end{center}
\vspace{-5mm}
\end{figure}

To evaluate the presented framework, a non-trivial SoC design meeting the following requirements is desired.
\begin{itemize}[leftmargin=*]
\setlength{\itemsep}{0pt}
\item It implements sophisticated system flows.   
\item Flows are well documented.
\item On-chip communication fabric is concurrent, and can support multiple active flow executions in parallel.
\end{itemize}
However, to the best of our knowledge, we cannot find an open-source design that meets the above requirements.  Therefore, we developed a multi-core SoC prototype, as shown in Figure~\ref{rtlstruc}, to evaluate the proposed methods.   Even though this model is simple compared to real SoC designs, it is much more sophisticated than the gate-level benchmark suites typically considered as targets for post-silicon debug research.  In particular,  while simplified, the system is representative of industrial complexity.  All the flows are abstracted from real industrial protocols.  Note that a typical industrial subsystem includes between $8$ and $32$ flows \cite{Talupur:2015:fmcad}.   We implement a total of $16$ system-level flows, including cache coherence, power management, downstream read/write protocols for CPUs, upstream read/write for the peripheral blocks, \emph{etc}.  A different paper~\cite{amrein2015system} details some of the flows.

We implemented this SoC as  a cycle- and pin-accurate RTL model written in VHDL.  The above system-level protocols are supported by block-level protocols based on the ARM {AXI4-lite~\cite{AXI4}}.  A total of $32$ monitors are inserted into this model guided by the implemented flows.
Since the proposed framework is to support communication-centric debug, the focus of this model is the implementation of system flows for on-chip interconnect.   The interconnect is a switch-based network, that allows multiple flow instances in-flight simultanenously.   The CPUs are treated as test generators where software programs are simulated in VHDL.  Therefore, there is no instruction cache.  Blocks GFX, PMU, Audio, \emph{etc}, are also described as abstract models that generate events to initiate flows or to respond incoming requests. 

\subsection{Experimental Results}
\label{sec:exp-results}

The prototype is simulated in a random test environment where CPUs and three other peripheral blocks are programmed to randomly select a flow to initiate with a delay between 1 to 10 cycles.  
%The contents of {\tt Cmd} and {\tt Addr} of the start events that initiate flows are set randomly.  
Each of these blocks activates {100} flow instances, and a total of $500$ flow instances are activated during the entire simulation.

% To illustrate different ideas presented in the previous section, two different test configurations are implemented:
% \begin{itemize}
% \item[S1] 
% To simulate the real communication traffic, each CPU issues an request after a random delay of 0 to 10 clock cycles when it receives the response for the previous request. While the peripherals issue request without waiting for responses such to maximize the communication activities density.
% % This experimental setup is used for experimenting the effects of buffer sizes, debug interface sizes, which flows to observe or links to observe, and critical ranking about how they change the trace analysis result regarding numbers of dropped events and observed flows.
% \item[S2]
% For the experiment regarding the impact of communication traffic density on the analysis result, the peripherals are reconfigured to wait for a response before issuing a new request such that the numbers of active flow at each clock cycle can be controlled by the idle cycles between each flow instance.
% \end{itemize}

% \subsection{Data}

% \begin{table}[tb]
% \centering{
% %  \resizebox{.5\textwidth}{!}{
%  \begin{tabular}{ |c|c|c|c|c|c|} 
%  \hline
% Scope& EDR&FIC&CEC\\
%   \hline
% All&1352/3154&470/500&101/500\\
%   \hline  
% CPU &302/1286&172/200&103/200\\
%   \hline
% CPU0 &0/626&100/100&100/100\\
% \hline
%  \end{tabular}
% % }
%  }
% \caption{Results with different flows under observation. All the queues is 8.}
% \label{table:scope}
% \vspace{-2mm}
% \end{table}

\begin{table}[tb]
\centering{
%  \resizebox{.5\textwidth}{!}{
 \begin{tabular}{ |c|c|c|} 
 \hline
Scope & FIC & CEC\\
  \hline
All &   470/500~(0.94) & 101/500~(0.202)\\
  \hline  
CPU     & 192/200~(0.96) & 103/200~(0.515)\\
  \hline
CPU0 & 100/100~(1) & 100/100~(1)\\
\hline
 \end{tabular}
% }
 }
\caption{Results with different flows under observation. The capacity of all the queues is 8.}
\label{table:scope}
\vspace{-2mm}
\end{table}

% \begin{table}[tb]
% \centering{
%  \resizebox{.5\textwidth}{!}{
%  \begin{tabular}{ |c|c|c|c|c|c|c|c|c|c|c|c|} 
%  \hline
% Scope&EDC &TEC&SEC&EEC&0\_rd&0\_wt&1\_rd&1\_wt&Audio&GFX&USB\\
%   \hline
% All&1874&1280&221&304&18&16&24&23&26&27&22\\
% \hline
% CPU &459&955&125&182&24&20&33&36&0&0&0\\
% \hline
% CPU0 &0&680&100&100&47&53&0&0&0&0&0\\
%  \hline
% \end{tabular}
%  }
%  }
% \caption{Analysis result with different scopes when no critical mode is implemented and the FIFO size is 16.}
% \label{table:scope}
% \end{table}

% \begin{table}[tb]
% \centering{
% %  \resizebox{.5\textwidth}{!}{
%  \begin{tabular}{ |c|c|c|c|c|c|} 
%  \hline
%   Delay & EDR&FIC&CEC\\
%   \hline
%   1-10&1352/3154&470/500&101/500\\
%   \hline  
%   11-20& 1277/3154 &487/500  &308/500 \\
%   \hline
% 21-30& 959/3154& 498/500&415/500 \\
% \hline
%  \end{tabular}
% % }
%  }
% \caption{Results with different delays between initiations of consecutive flow instances. Priority output mode is not used, and the size of all the queues is 8.}
% \label{table:traffic}
% \end{table}

% \begin{figure}
% \centering
% \includegraphics[width=.5\textwidth]{figures/nocriti.png}
% \caption{Results on the observed events, and from the trace analysis with different queue sizes.  Priority output is not used. \textcolor{red}{should i put the number as ratio?}}
% \label{fig:nocriti}
% \end{figure}

% \emph{In each experiment, only an unique flow execution scenario is derived.}  This result shows the effectiveness of including micro-architecture information in events on the accuracy of the trace analysis. 
Table \ref{table:scope} shows the experimental results when different flows are observed.
Under the columns FIC and CEC, the numbers $A/B$ represent the ratios as defined in Section~\ref{sec:observe-config}. The equivalent fractional numbers for $A/B$ are enclosed in parentheses. 
The second row shows the results when all $500$ flow instances are observed.
The third row shows the results from observing only flows initiated by CPU0 or CPU1.   The last row shows the results from observing only the flows initiated by CPU0. 
From the table, it can be seen that as the number of flows under observation decreases, the number of events that need to be observed for FIC and CEC decreases.  This leads to reduced losses of events, which is reflected in the higher FIC and CEC values from row two to row four.   Particularly, in the last row, all executed instances of the flows under observation can be precisely inferred.
\begin{table}[tb]
\centering{
%  \resizebox{.5\textwidth}{!}{
 \begin{tabular}{ |c|c|c|} 
 \hline
FIFO\_SIZE &  FIC & CEC \\
\hline
8& 470/500~(0.94) & 101/500~(0.202) \\ %&1352/3154\\
\hline  
16& 490/500~(0.98) & 178/500~(0.356) \\ %&1104/3154 \\
\hline
32& 499/500~(0.998) & 307/500~(0.614) \\ %& 768/3154 \\
% \hline
% 64&500/500&455/500&341/3154 \\
% \hline
% 128&500/500&500/500&42/3154 \\
\hline
\end{tabular}
}
\caption{Impacts of different capacities of event queues.}
\vspace{-4mm}
\label{table:comp1}
\end{table}

In the second experiment, the impacts of different queue capacities are evaluated.  All flows and all flow events are observed.  Event dropping happens only when a queue becomes full.  In this experiments, the capacities of event queues are increased gradually.  This increase in queue capacity can simulate the situation where the number of links to observe is reduced and the queues of the non-observable links are re-allocated to the queues of links under observation.  The results are shown in Table~\ref{table:comp1}.  It is noticeable that an increase in the size of queues leads to a significant improvement in FIC and CEC values of observed traces.  These results validate the hypothesis described in Section~\ref{sec:priority-output}.
% Then the monitoring infrastructure is configured with same setting, but its output ports is increased from 1 to 2 for the second experiment. In the third experiment, the funnel has the same setting as the first experiment, but its output port speed is doubled. 

\begin{table}[tb]
\begin{center}
\resizebox{.5\textwidth}{!}{
 \begin{tabular}{ |l|l|l|c|} 
  \hline
  Selection method &\#\_Links& FIC & CEC \\ %&EDR\\
  \hline
  NO-SELECTION & 32(total)& 470/500~(0.94) & 101/500~(0.202) \\ %&1352/3154\\
   \hline
  SEL1 (FIC)  &  8& 500/500~(1) & 0/500~(0)    \\ %&2057/3154\\
  \hline
  SEL2 (CEC)  & 16&  441/500~(0.882) & 200/500~(0.4)    \\ %&2057/3154\\
  \hline
  SEL3 \cite{amrein2015system} & 16&466/500~(0.932) &   0/500~(0)   \\ %&1528/3154 \\
\hline
 \end{tabular}
 }
\end{center}
\vspace{-2.5mm}
\caption{Comparisons of different event selection methods}
\label{table:comp}
\vspace{-4mm}
\end{table}
Table \ref{table:comp} compares the result of the proposed flow event selection algorithm with another system level flow guided selection approach proposed in \cite{amrein2015system}. We generate two sets of selections guided by $FIC$ and $CEC$ individually ($SEL1$ and $SEL2$) and compare their results with selection $SEL3$ generated by  \cite{amrein2015system}.
And the results are evaluated by the previously mentioned two metrics and the numbers of links used.
The second row $NO$-$SELECTION$ represents the analysis result when no selection method is applied, and all flow events on the 32 links are observed.

The $SEL1$ on the third row is generated by our method that is primarily guided by the $FIC$. In the beginning, several sets of flow events with the optimal $FIC$ effects (covering all flows) are generated, and within them we select the set with the minimal numbers of links required, that is $8$ in this situation. The result shows that the $FIC$ is improved to its maximum value $1$. While on the other hand $CEC$ is reduced to $0$. This is expected as $CEC$ is not considered for this selection. Then for the second selection $SEL2$, we considers the $CEC$ only and its result is shown in the fourth row. It first selects all 32 initiating and terminating events of all flows to enhance $CEC$. Because several flow events are transferred on the same link (for example, both  {\tt (CPU\_0:Cache\_0:wr\_req)} and \ {\tt (CPU\_0:Cache\_0:rd\_req)} are transferred on the same link), these events takes 12 links in total.
Moreover, $SEL2$ selects four flow events including $t_2$ and $t_4$ that indicates the path of a flow, as discussed in Section~\ref{sec:event_selection}, occupying 4 links. 
Compared to the $NO$-$SELECTION$, The result of $SEL2$ takes only 16 links, and it shows significant improvement on the $CEC$ (almost double) as more start and end events are observed. Consequently, more details regarding the interleaving relationship of fired flow instances are revealed in $SEL2$. This improvement comes in the cost of less flow instances being observed, shown by the decrease in $FIC$ value.

The fifth row $SEL3$ shows the result of using flow selection method from \cite{amrein2015system}. 
\cite{amrein2015system} proposes to rank each flow event by their Frequency Coverage ($FC$) value in descending order, and apply greedy algorithm to select the optimal set of flow events.
The $FC$ considers the fact that some flow events are shared by multiple flows thus are always preferred.
The concept of $FC$ is very similar with $FIC$ as it enforces the maximum number of flows being covered for a set of flow events. However, the $FC$ in \cite{amrein2015system} is different in two aspects: (1) it does not consider the number of links used for such selection. Consequently, the capacity improvement may not be as significant as our method where we always select the set with the minimal number of links; (2) Because $FC$ considers each flow event individually, it is possible the selected combination of events achieves high coverage only on certain flows. While our method ensures that the selected flow event set covers all flows.

We applied the algorithm in \cite{amrein2015system} and selected top $16$ (half of the total link number) flow events with the highest $FC$. 
The result of $SEL3$, however, does not show any improvement compared to $NO$-$SELECTION$, the $FIC$ actually reduced from $470$ to $466$ due to the two factors mentioned above. 
It is also to be noted 
% that the highly shared events are likely to be fired more frequently compared to others as it can be invoked by multiple flows. In order to reduce the drop rate of such events, the available queue capacities is consequently higher. 
for $SEL3$, each of the selected flow occupies one link individually, taking 16 links in total, that is double of $SEL1$. As a result, the $FIC$ of $SEL3$ is comparably worse than both $SEL1$ and $NO$-$SELECTION$.
On the other hand, none of the initiation and termination events are selected due to their uniqueness to their belonging flows (not shared by any other flows), leading to $0$ $CEC$. This is expected as this algorithm does not consider special meanings of such critical events.

%We implement and synthesis 3 types of monitors for AXI read, AXI write and customized protocols using an commercial tool. Each of them with targeted communication protocols built inside.

% \begin{table}[tb]
%  %\resizebox{.4\textwidth}{!}{\begin{minipage}{.4\textwidth}
% \resizebox{.47\textwidth}{!}{
%  \begin{tabular}{ |c|c|c|c|c|c| } 
%  \hline
%  &Cells& LUTs & FFs &Muxs& BRAM\\
%  \hline
% Design &	50973 & 18612 & 24121 &	3473 &	1 \\
%  \hline
% +Debug &+315	 &+119  &+822	 &	-16 &+33	\\
%  \hline
% \end{tabular}
% }
% \caption{Area overhead of the monitoring infrastructure.}
% \label{table:synthesis}
% %\end{minipage}}
% \end{table}

\vspace{3pt}
\noindent{\bf Area overhead.~} The SoC model is synthesized to the Xilinx Zynq FPGA xc7z020ckg484-1.  
% The synthesis results are shown in Table~\ref{table:synthesis}. 
The area overhead of the debug infrastructure is measured by the additional FPGA resources including LUTs, FFs, block RAMs (BRAMs), etc. From the obtained result, the demand on logic resources is small to implement the monitoring infrastructure; On tops of the original design area, this new implementation requires additional $315$ LUTs, $119$ FFs, and $822$ Muxs, that are $0.6\%$, $0.6\%$ and $3\%$ of the original design area. 
% From the table, other than the big jump in BRAM usage, the demand on logic resources is small to implement the monitoring infrastructure.  Due to the specific feature of the FPGA synthesis tool, each queue in Figure~\ref{fig:funnel2} is implemented with one BRAM.  Since the size of BRAM is fixed, the queue size is determined by its width. For example, the queues used in the above experiments are 32-bit wide, thus a 36Kb BRAM allows the $1024$ queue depth even though the queue depths used in the above experiments are much smaller. For future work, ASIC synthesis will be performed so that the area overhead can be evaluated more precisely.

% The synthesis results are obtained from a design prototype without the actual implementations of CPUs and peripheral blocks. Otherwise, the area overhead for the monitoring infrastructure would be much smaller.  Furthermore, modern SoCs are often instrumented with on-chip trace buffers.  The memory footprint would be reduced further if those trace buffers can be reused for the presented framework.

%% file: 6-rel-work.tex
\section{Related Work}

Gharehbaghi and Fujita \cite{5423891,5413157,6187568} describe transaction-level debug with an on-chip instrumentation that allows transaction level message abstraction using formal specifications of the bus communication protocols. 
%  They propose to consider only start and end of transactions to build the transaction sequence from the signal events; As the result, the area overhead is very low since the instrumentation does include the full communication protocols. 
%The authors propose an innovative encoding technique that can reduce the address bits to two bits containing three different states % : SAME,SEQ,OTHER. This state efficiently  to indicate relationships between two consecutive transactions.
% where $SAME$ indicate same slave address as previous translation slave address; $SEQ$ specifies one word difference with the previous transaction address; and $OTHERS$ specifies any other situation.
% The authors also present a post analysis method targeted at recognizing potential erroneous patterns such as deadlock and racing condition.  Despite its low area overhead, this methods suffers from inability of detecting implementation errors that are are not observed. 
Their approach does not check system-level protocols as it only focus on communication protocols among component interfaces.
\cite{6386594} presents another on-chip instrumentation BiPeD that learns communication interface' protocol during pre-silicon stage, and reconfigure its detection hardware to check the learned protocols during the post-silicon validation. 
% Once a bug is detect, recent history of observed activity is transferred off-chip for analysis. 
% With the recent history of the protocol level activity related to detected bug, BiPed provides rich set of debugging information including bug location, occurring time, and related critical signals, and et al.
% While BiPed is effective towards detecting and locating many hardware bugs, 
% % the quality of the learned protocols depends heavily on pre-silicon
% % , where its completeness and correctness can not be guaranteed; 
% % Moreover,
% the circular buffer implemented for each communication interface introduces large area overhead.  
Similar to~\cite{5423891,5413157,6187568}, interpreting the observed traces are not done at the system level.
\cite{chatterjee2012checking} presents a processor verification framework on an acceleration platform.  Although both \cite{chatterjee2012checking} and this paper consider interpretation of low level behavior on hardware signals at a higher level, 
%\cite{chatterjee2012checking} does that \emph{wrt} a golden processor model while this work focuses on executions of system-level protocols.  
\cite{chatterjee2012checking} considers verification of individual processors, while this work targets post-silicon debug of SoC designs where processors are just some IP blocks.  Since the work in~\cite{chatterjee2012checking} is developed for an acceleration platform, its monitoring infrastructure does not consider area restriction of the monitors.  On the other hand, area overhead is a top restriction in our monitor design.  Furthermore, the design under verification can be slowed down to match the speed of the monitors to make sure that all observed events can be outputted~\cite{chatterjee2012checking}.  However, on-chip debug infrastructure cannot interfere with normal chip operations in the post-silicon environment where this work is positioned.

Previous work on trace signal selection such as~\cite{mishra2011vlsi} is typically applied to gate level design models, and the quality of the results is evaluated by the commonly used state restoration ratio.  However, it is difficult to scale those methods to large and complex SoC designs. More importantly, signals selected at the gate level are often irrelevant to system-level functionalities.  There is an attempt to raise the abstraction level for trace signal selection to the register transfer level~(RTL) guided by assertions~\cite{Ma2015ICCAD}, however that work does not consider system level functionalities either.  On the other hand, our framework considers selection of communication events guided by system-level protocols, instead of raw hardware signals.  
\cite{pal2018application} proposed an similar concept where signal selection is conducted on system-level communication protocols.
 While \cite{pal2018application} shows significant improvement compared to other low level based signal selection algorithms, its effect is mitigated by the resource limitation of the on-chip debug infrastructure. 
The selecting scheme proposed in \cite{pal2018application} is guided by a metric that treat every communication protocol event equally. It failed to consider special events that are critical for an comprehensive interpretation of the system behavior. Furthermore, the algorithm in \cite{pal2018application} requires the interleaving graph of all flow instances to generate a set of selection. Obtaining such interleaving graph requires the type and number of instances of each flow. 
Such requirement is impractical as itself is one of the critical information needed to be extracted during post silicon validation. 
% Moreover, even if such information is obtained, there could be more than thousands of instances of each flow, leading to exponential growth in the size of possible interleavings that is too expensive to analyze.

%In~\cite{amrein2015system}, a system level protocol guided approach similar to this work is proposed.  However, the selection techniques developed in~\cite{amrein2015system} are simple and irrelevant to understanding silicon traces at the system level, and the evaluation was performed on an abstract transaction level model.

%\input{related-work-monitor}

% Lamport in \cite{Lamport:1978:TCO:359545.359563} proposes an interesting idea to synchronize a system of logical clocks to obtain a global event order. 
% Where the partial ordering "happening before" is extended to a somewhat arbitrary global ordering. 
% And this work is further extended by Gharehbaghi and Fujita in \cite{5770738}. Where they applies Lamport's algorithm into a Network-on-chip for post-silicon validation. To improve the accuracy of extracted order, the local partial ordering is enriched by considering some information from the neighboring tiles.